\documentclass[aps,prb,groupedaddress,twocolumn,showpacs]{revtex4}
\usepackage{graphicx,psfrag,amsmath,calc}

\begin{document}

\title{F-wave versus P-wave Superconductivity
\\
in Organic Conductors}

\author{R. W. Cherng$^\dagger$}
\author{C. A. R. S\'{a} de Melo}
\affiliation{School of Physics, Georgia Institute of Technology,
             Atlanta Georgia 30332}

\date{\today}

\begin{abstract}
Current experimental results suggest that some organic 
quasi-one-dimensional superconductors exhibit
triplet pairing symmetry. 
Thus, we discuss several potential triplet order parameters for
the superconducting state of these systems
within the functional integral formulation.
We compare weak spin-orbit coupling 
$f_{xyz}$, $p_x$, $p_y$ and $p_z$ symmetries via several
thermodynamic quantities. 
For each symmetry, we analyse the 
temperature dependences of the order parameter,
condensation energy, specific heat, and superfluid density tensor.
\end{abstract}

\maketitle

Recent NMR experiments in the organic conductor (Bechgaard salt) 
${\rm (TMTSF)_2 PF_6}
$~\cite{lee-00,lee-02} in combination with earlier upper critical
field $(H_{c_2})$ measurements~\cite{lee-97} indicate that this
system may be an unconventional triplet superconductor
at low temperature $(T)$. 
Lee {\it et. al.}~\cite{lee-00,lee-02} observed that
there is {\bf no} $^{77} {\rm Se}$ Knight shift
for fields ${\bf H} \parallel {\bf b}^{\prime}$ 
$(P \approx 6~{\rm kbar})$ and 
${\bf H} \parallel {\bf a}$ 
$(P \approx 7~{\rm kbar})$.
This indicates that the spin susceptibilities 
in the superconducting state 
are $\chi_a \approx \chi_N$, and $\chi_{b^\prime} \approx \chi_N$, 
where $\chi_N$ is the normal state susceptibility.
Their measurements suggest that
the ${\bf d}$-vector order parameter for triplet superconductivity
is pointing along the ${\bf c}^*$ axis. Unfortunately,
the spin susceptibility in the superconducting state 
for fields along the ${\bf c}^*$ axis could not 
be measured because the upper critical field $(H_{c_2}^{c^*})$
was exceeded before a signal could be detected.  
Since the low $T$ dependence
of $\chi_{c^*} (T)$ is not accessible experimentally, it may not
be possible to use this technique to decipher the node 
structure of the triplet order parameter.
The symmetry of the order parameter in the sister compound
${\rm (TMTSF)_2 ClO_4}$ was preliminarily explored in the
thermal conductivity experiments of 
Belin and Behnia~\cite{belin-97}
(BB).  They indicated that their data was inconsistent
with the existence of gap nodes at the Fermi surface as suggested
by Takigawa {\it et. al.}~\cite{takigawa-87}. 
These results combined with upper critical field 
measurements~\cite{lee-95,lee-98}
seem to suggest a fully gapped triplet state in
${\rm (TMTSF)_2 ClO_4}$. However, detailed  
$^{77} {\rm Se}$ NMR experiments seem to be 
lacking for ${\rm (TMTSF)_2 ClO_4}$.

These experiments were inspired by early
suggestions of triplet superconductivity
in the Bechgaard 
salts~\cite{abrikosov-83,lebed-86,dupuis-93,sademelo-96},
but following these new experimental developments 
theoretical efforts 
intensified~\cite{lebed-00,shimahara-00,kuroki-01,duncan-01,duncan-02}.
Lebed, Machida, and Ozaki (LMO)~\cite{lebed-00} 
proposed a ``p-wave'' triplet order parameter for 
${\rm (TMTSF)_2 PF_6}$,
where the ${\bf d}$-vector had a strong
component along the ${\bf a}$ direction, thus producing a strongly anisotropic spin susceptibility 
with $\chi_{a} \ll \chi_N$ and $\chi_{b^\prime} \approx \chi_N$.
A fully gapped singlet ``d-wave'' order parameter
for ${\rm (TMTSF)_2 ClO_4}$ was 
proposed by Shimahara~\cite{shimahara-00},
while gapless triplet ``f-wave'' superconductivity 
for ${\rm (TMTSF)_2 PF_6}$ was proposed by 
Kuroki, Arita, and Aoki (KAA)~\cite{kuroki-01}.
Duncan, Vaccarella and S\'a de Melo (DVS)~\cite{duncan-01,duncan-02}
performed a detailed group theoretical analysis and 
suggested that a weak spin-orbit fully gapped 
triplet ``$p_x$-wave'' order parameter, where
$\chi_{a} \approx \chi_N$ and 
$\chi_{b^\prime} \approx \chi_N$~\cite{sademelo-98,sademelo-99},
would be a good candidate for superconductivity 
in Bechgaard salts.

In this paper, we consider only triplet states 
corresponding to weak spin-orbit coupling, since
in ${\rm (TMTSF)_2 X}$ the heaviest element is
$^{77} Se$. 
Within the orthorhombic $(D_{2h})$ group,
this limits the number of possibilities
for the order parameter~\cite{duncan-01,duncan-02}
to four symmetries, 
$p_x$, $p_y$, $p_z$, $f_{xyz}$.
For each one of these symmetries, we calculate the 
temperature dependences of the 
order parameter, condensation energy, entropy, specific heat, 
and superfluid density tensor.

We study single band quasi-one-dimensional 
systems in an orthorhombic
lattice with 
dispersion relation
%
%%
%%\begin{equation}
%%\label{eqn:dispersion}
$$
\epsilon_{\bf k} = -|t_x| \cos(k_x a_x) - 
|t_y| \cos(k_y a_y) - |t_z| \cos(k_z a_z),
$$
%%\end{equation}
%%
%
where $|{t_x}| \gg |{t_y}| \gg |{t_z}|$.
Furthermore, $a_x$, $a_y$ and $a_z$ in our notation correspond to the
unit cell lengths along the crystallographic directions
${\bf a}$, ${\bf b}^{\prime}$, and ${\bf c}^*$ respectively.
We work with the Hamiltonian
$
H = H_{kin} + H_{int},
$
where the kinetic energy part is 
$
H_{kin} = \sum_{{\bf k},\alpha} \xi_{\bf k}
\psi_{{\bf k}, \alpha}^{\dagger} \psi_{{\bf k}, \alpha},
$
with $\xi_{\bf k} = \epsilon_{\bf k} - \mu$ 
and the interaction part is
\begin{equation}
\label{eqn:hint}
H_{int} = 
\dfrac{1}{2} \sum_{{\bf k} {\bf k^{\prime}} {\bf q}} 
\sum_{\alpha \beta \gamma \delta}
V_{\alpha \beta \gamma \delta} ({\bf k}, {\bf k^{\prime}})
b_{\alpha \beta}^{\dagger} ({\bf k}, {\bf q})
b_{\gamma \delta} ({\bf k^{\prime}}, {\bf q})
\end{equation}
with
$
b_{\alpha \beta}^{\dagger} ({\bf k}, {\bf q}) = 
\psi_{-{\bf k} + {\bf q}/2, \alpha}^{\dagger}
\psi_{{\bf k} + {\bf q}/2, \beta}^{\dagger},
$
where $\alpha$, $\beta$, $\gamma$ and $\delta$ 
are spin indices and 
${\bf k}$, ${\bf k}^{\prime}$ and ${\bf q}$ represent 
linear momenta. We use units where $\hbar = k_B = 1$. 

In the case of weak spin-orbit coupling
and triplet pairing, the model interaction tensor can be chosen to be
\begin{equation}
V_{\alpha \beta \gamma \delta} ({\bf k}, {\bf k^{\prime}})
= V_{\Gamma} 
h_{\Gamma} ({\bf k}, {\bf k^{\prime}})
\phi_{\Gamma} ({\bf k}) \phi^{*}_{\Gamma} ({\bf k^{\prime}})
\Gamma_{\alpha \beta \gamma \delta},
\end{equation}
where 
$\Gamma_{\alpha \beta \gamma \delta} = {\bf v}_{\alpha \beta} \cdot
{\bf v}_{\gamma \delta}^{\dagger}/2$ with 
${\rm v}_{\alpha \beta} = (i\sigma \sigma_y)_{\alpha \beta}$. 
$V_{\Gamma}$ is a prefactor with dimensions of energy which characterizes 
a given symmetry. Furthermore, the term 
$h_{\Gamma} ({\bf k}, {\bf k^{\prime}}) 
\phi_{\Gamma} ({\bf k}) \phi^{*}_{\Gamma} ({\bf k^{\prime}})$ 
contains the momentum and symmetry dependence of the interaction 
of the irreducible representation $\Gamma$ with basis function
$\phi_{\Gamma} ({\bf k})$ and 
$\phi_{\Gamma}^{*} ({\bf k^\prime})$ 
representative of the orthorhombic group ($D_{2h}$).  

We use the functional integration method
to write down the partition function
$Z = \int {\cal D} \left[\psi^\dagger, \psi \right] {\rm exp} 
\left[ S \right] $
where $S = \int d\tau
\left[
\sum_{{\bf k}, \alpha} \psi_{{\bf k}, \alpha}^{\dagger} (\tau) 
(-\partial_\tau ) 
\psi_{{\bf k}, \alpha} (\tau) 
- H (\psi^{\dagger}, \psi) 
\right].$
We treat $H_{int}$ in 
the zero center of mass momentum $({\bf q} = {\bf 0})$  pairing approximation.
Furthermore, since we are mostly interested
in symmetry aspects we take 
$h_{\Gamma} ({\bf k}, {\bf k^{\prime}}) = 1 $.
We introduce the Gaussian integral
$I_G = \int d \left[ {\cal D}_{i}^{\dagger}, {\cal D}_{i} \right]
{\rm exp} \left[ - Q_{G} \right]$, 
where 
$Q_{G} = \int d\tau \sum_{ {\bf k} {\bf k^\prime} }
{\cal D}^{\dagger}_{i} ({\bf k},\tau) 
{\cal D}_{i} ({\bf k^\prime},\tau)/V_{\Gamma}$. 
The shift transformation
$ 
{\cal D}^{\dagger}_{i} ({\bf k}, \tau) \to  
{\cal D}^{\dagger}_{i} ({\bf k}, \tau) 
+ V_{\Gamma} {\rm v}_{\alpha\beta, i} \phi_{\Gamma} ({\bf k})
\psi^{\dagger}_{ {\bf -k}, \alpha } 
\psi^{\dagger}_{ {\bf k}, \beta }
$
eliminates $H_{int}$
and integration over the fermionic degrees of freedom
results in the effective action
\begin{equation}
\label{eqn:eff-action}
S_{\rm eff} = 
- Q_{G}
- Tr \ln \left[ {\bf M}/2 \right],
\end{equation}
where ${\bf M}$ is a $2 \times 2$ block diagonal matrix
of the form
$$
\left(
\begin{array}{cc}
\left[ \partial_{\tau} + \xi_{\bf k} \right] \delta_{\alpha \beta} & 
\sum_{\bf k^\prime} {\cal D}_{i} ({\bf k^\prime}, \tau) 
{\rm v}_{\beta \alpha, i} \phi_{\Gamma} ({\bf k}) \\
\sum_{\bf k^\prime} {\cal D}_{i}^{\dagger} ({\bf k^\prime}, \tau) 
{\rm v}_{\alpha \beta, i}^{\dagger}
\phi_{\Gamma}^{*} ({\bf k}) & 
\left[ - \partial_{\tau} - \xi_{\bf k} \right] \delta_{\alpha \beta} \\ 
\end{array}
\right). 
$$
Thus, $Z = \int d  \left[{\cal D}_{i}^\dagger, {\cal D}_{i} \right] {\rm exp} 
\left[ S_{\rm eff} \right] $.
At the saddle point approximation 
${\cal D}_{i} ({\bf k}, \tau)$ is taken to be 
$\tau$ independent, and 
$
\sum_{\bf k^\prime} {\cal D}_{i} ({\bf k^\prime}) 
\equiv 
\hat{\eta}_i \Delta_{\Gamma}.
$
The order parameter equation is obtained from the 
stationary condition $\delta S_{\rm eff}^{(0)}/\delta {\cal D}_{i}^{\dagger} 
= 0$ leading to
\begin{equation}
\label{eqn:order-parameter}
1 = - \sum_{\bf k} V_{\Gamma} |\phi_{\Gamma} ({\bf k})|^2 
{\rm tanh} (E_{\bf k}/2T) / 2 E_{\bf k},
\end{equation}
where 
$
E_{\bf k} = 
\sqrt{ \xi_{\bf k}^2 + |\Delta_{\Gamma}|^2 |\phi_{\Gamma} ({\bf k})|^2}.
$
The number equation is obtained from 
$N = - \partial \Omega_0/\partial \mu$, 
where  $\Omega_0  = - T S_{\rm eff}^{(0)}$ is 
the saddle point thermodynamic potential,
and results in
\begin{equation}
\label{eqn:number}
N = \sum_{\bf k} n_{\bf k},
\end{equation}
where $
n_{\bf k} = 
\left[
1 - \xi_{\bf k} {\rm tanh} (E_{\bf k}/ 2T)/E_{\bf k} 
\right]
$
is the momentum distribution.
These two equations must be solved self-consistently, and 
quite generally they are correct even in the strong coupling (or low density)
regime provided that $T \ll T_c$.
Corrections to $\Omega$ and $N$ 
can be obtained by considering
Gaussian fluctuations. Writing ${\cal D}_{i} ({\bf k}, \tau ) = 
{\cal D}_{i} ({\bf k}) +
\delta {\cal D}_{i} ({\bf k}, \tau )$ and expanding 
$Tr \ln \left[ {\bf M} \right]$ 
to quadratic order in $\delta {\cal D}_{i} ({\bf k}, \tau )$ results
in the effective action  
$
S_{\rm eff} = 
S_{\rm eff}^{(0)} - Tr \left[{\bf M}^{-1} {\bf U}\right]^2 /2,
$
where ${\bf U}$ is a $ 2 \times 2$ block matrix that contains only 
off-diagonal elements, with 
$U_{12} = 
\sum_{{\bf k^\prime}} \delta {\cal D}_{i} ({\bf k^\prime}, \tau ) 
v_{\beta \alpha, i} 
\phi_{\Gamma} ({\bf k})$ and $U_{21} = U_{12}^{\dagger}$. 
These corrections are important for small carrier density
(or $T_c \sim E_F$), 
however, in the BCS limit of high carrier density (or $T_c \ll E_F$), 
discussed here, these corrections are negligible for $N$,
but are important for $\Omega$ only when $T \ll T_c$.
For Bechgaard salts, $T_c \approx 1.5~{\rm K}$ 
and $E_F \approx 3,071~{\rm K}$ (with respect
to the bottom of the band).
\begin{figure}
\begin{center}
\(
  \begin{array}{c}
   \multicolumn{1}{l}{\hspace{-0.2cm}\mbox{\bf (a)}} \\
	\psfrag{T (K)}{$T\ (K)$}
	\psfrag{Dg (K)}{$\Delta_{\Gamma}\ (K)$}
     \includegraphics[width=7.0cm]{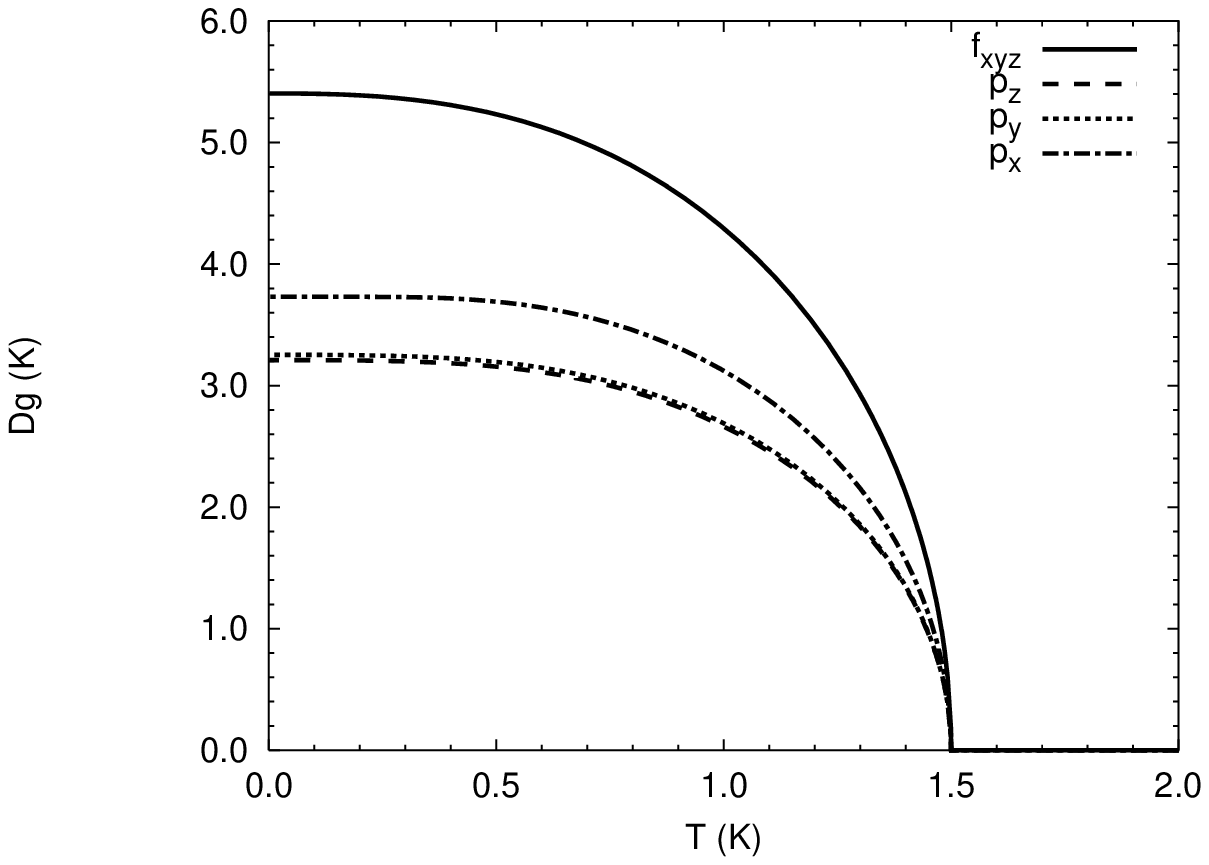} \\
   \multicolumn{1}{l}{\hspace{-0.2cm}\mbox{\bf (b)}} \\
     \psfrag{T (K)}{$T\ (K)$}
	\psfrag{ug (K)}{$\mu_{\Gamma}\ (K)$}
\includegraphics[width=7.0cm]{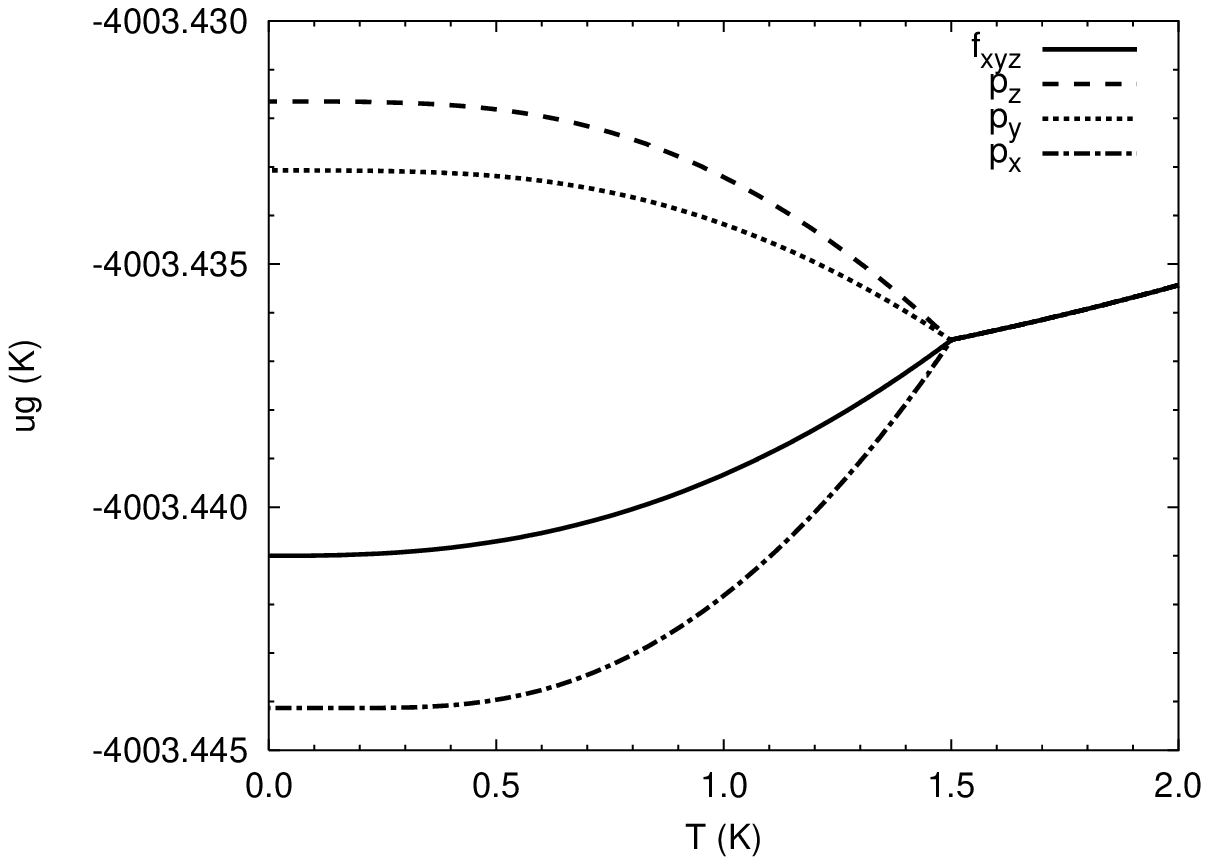}
\end{array}\)
\end{center}
\caption{
Temperature dependence 
for (a) $\Delta_{\Gamma}$ and (b) $\mu_{\Gamma}$ in thermal
units $(K)$. Notice that $\mu_{\Gamma}$ is nearly $T$ independent.
}
\label{fig:delta-mu}
\end{figure}
The vector ${\cal D}_i $ is related to the standard ${\bf d}$-vector
via the relation 
$
{d_i} ({\bf k}) = 
\sum_{\bf k^\prime} {\cal D}_i ({\bf k^\prime}) \phi_{\Gamma} ({\bf k}).
$
In the $D_{2h}$ point group all representations are one dimensional 
and non-degenerate~\cite{duncan-01,duncan-02}, which means that
the ${\bf d}$-vector in momentum space for unitary triplet
states in the weak spin-orbit coupling limit
is characterized by one of the four states:
(1) $^3 A_{1u} (a)$, 
with  
${\bf d} ({\bf k}) = {\hat \eta} \Delta_{f_{xyz}} X Y Z$ 
(``$f_{xyz}$'' state);
(2) $^3 B_{1u} (a)$, 
with  
${\bf d} ({\bf k}) = {\hat \eta} \Delta_{p_z} Z$ (``$p_{z}$'' state);
(3) $^3 B_{2u} (a)$, 
with  
${\bf d} ({\bf k}) = {\hat \eta} \Delta_{p_y} Y$ (``$p_{y}$'' state);
(4) $^3 B_{3u} (a)$, 
with  
${\bf d} ({\bf k}) = {\hat \eta} \Delta_{p_x} X$ (``$p_{x}$'' state).
Since, the Fermi surface touches
the Brillouin zone boundaries the functions 
$X$, $Y$, and $Z$ need to be periodic and can 
be chosen to be $X = \sin{(k_{x} a_x)}$, 
$Y = \sin{(k_{y} a_y)}$, 
and $Z = \sin{(k_{z} a_z)}$.
The unit vector $\hat \eta$ defines the direction 
of ${\bf d} ({\bf k})$. 
The $T$ dependence of $\Delta_{\Gamma}$ and $\mu_{\Gamma}$ 
for $f_{xyz}$, $p_x$, $p_y$ and $p_z$ symmetries 
are shown in Fig.~\ref{fig:delta-mu}.
The parameters used are $|t_x| = 5800~{\rm K}$, 
$|t_y| = 1226~{\rm K}$ and $|t_z| = 48~{\rm K}$, and
$N/N_{max} = 1/4$ (quarter-filling). The minimum value of the
dispersion $\epsilon ({\bf k})$ is 
$\epsilon_{min} = -7,074~{\rm K}$, while
the maximum is $\epsilon_{max} = +7,074~{\rm K}$. 
In order to make direct comparison between different symmetries
we choose $T_c = 1.5~{\rm K}$ for all symmetries.
This requirement forces the interaction strength 
prefactors to be
$V_{f_{xyz}} \approx  -12,577~{\rm K}$; 
$V_{p_{z}}   \approx  -2,910~{\rm K}$; 
$V_{p_{y}}   \approx  -3,025~{\rm K}$; 
$V_{p_{x}} \approx  -3,208~{\rm K}$. 
Notice in Fig.~\ref{fig:delta-mu}a that 
$\Delta_{f_{xyz}} > \Delta_{p_{x}} > 
\Delta_{p_{y}} > \Delta_{p_{z}}$
for all $T < T_c$, and that 
$\mu_{\Gamma}$ is largely
independent of $T$, however, the 
condition 
$
\mu_{p_{z}} > \mu_{p_{y}} > \mu_{f_{xyz}} > \mu_{p_x}
$ applies
for all $T < T_c$.
\begin{figure}
\begin{center}
  \psfrag{T (K)}{$T\ (K)$}
	\psfrag{DFg (K)}{$\Delta{\cal F}_{\Gamma}\ (K)$}
\includegraphics[width=7.0cm]{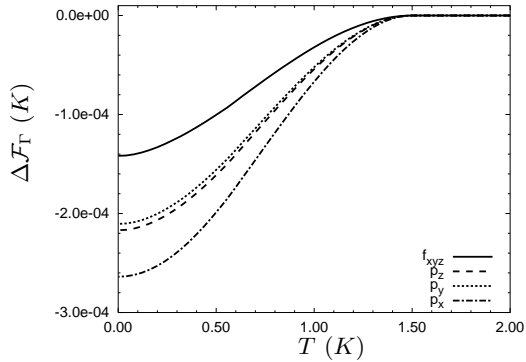} \\
\end{center}
\caption{
Temperature dependence of the condensation energy $\Delta {\cal {F}}_{\Gamma}$
in thermal units $(K)$.
\label{fig:condensation-energy}
}
\end{figure}
The relative stability of these phases
can be studied via the condensation energy 
$\Delta {\cal {F}}_{\Gamma} = {\cal{F}}_{\Gamma} (T) - {\cal{F}}_{N} (T) $,
where ${\cal{F}}_{\Gamma} (T)$ and ${\cal {F}}_{N} (T)$ are
the Helmholtz free energies of the superconducting state with
symmetry $\Gamma$ and of the normal state, respectively.
$\Delta {\cal{F}}_{\Gamma}$ is calculated using 
${\cal {F}} = \Omega + \mu N$. 
Fig.~\ref{fig:condensation-energy} shows 
$\Delta {\cal{F}}_{\Gamma}$ for the
$f_{xyz}$, $p_x$, $p_y$ and $p_z$ symmetries.
Notice that $\Delta {\cal{F}}_{\Gamma}$ 
are very small, are  expressed in thermal units,
and obey the relations
$
|\Delta {\cal {F}}_{p_x}| >
|\Delta {\cal {F}}_{p_z}| > |\Delta {\cal{F}}_{p_y}| >
|\Delta {\cal {F}}_{f_{xyz}}|
$
for $T < T_c$.
The $p_x$ symmetry which has a full gap, 
has also the largest (negative) $\Delta {\cal{F}}_{\Gamma}$. 
The values of $\Delta {\cal {F}}_{p_z}$ and $\Delta {\cal {F}}_{p_y}$ are
very close reflecting that the $p_z$ and $p_y$ symmetries
have lines of nodes at $Z = 0$ and $Y = 0$, respectively. 
Lastly, the $f_{xyz}$ symmetry has double zeros 
at $Z = 0$ {\bf and} $Y = 0$ and has line nodes at $Z = 0$ {\bf or} $Y = 0$, 
which are costly in condensation
energy. This confirms the general 
expectation that a fully gapped (nodeless) superconducting
phase ($p_x$) is more likely to win over 
competing phases which have nodes
($p_y$, $p_z$ and $f_{xyz}$).

\begin{figure}
\begin{center}
  \psfrag{T (K)}{$T\ (K)$}
	\psfrag{Cg (kb)}{$C_{\Gamma}\ (k_{B})$}
\includegraphics[width=7.0cm]{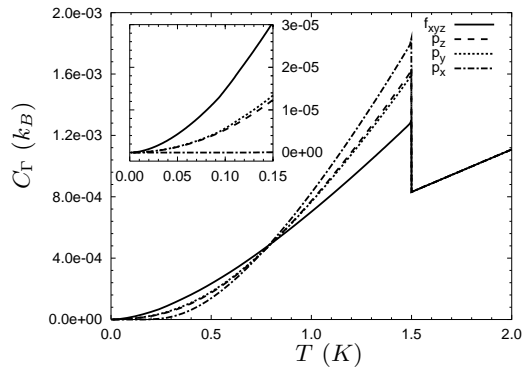} \\
\end{center}
\caption{
Temperature dependence of specific heat $C_{\Gamma}$ in units of $k_B$.
Inset shows low $T$ behavior.
\label{fig:specific-heat}
}
\end{figure}

Since $\Omega_0$ 
is only a function of 
$|{\bf d} ({\bf k})|^2$ (recall that 
$|{\bf d} ({\bf k})|^2 = |\Delta_\Gamma|^2 |\phi_{\Gamma} ({\bf k})|^2$)
the vector nature of ${\bf d} ({\bf k})$ does not appear explicitly in 
thermodynamic properties. However, the node structure of 
triplet order parameters can be probed 
by a specific heat measurement.
The specific heat 
$C_{\Gamma} = - T \partial^2 {\cal F}_{\Gamma}/\partial T^2$ is  
%
%%
%%\begin{equation}
%%\label{eqn:specific-heat}
$$
C_{\Gamma}  = {2 \over T^2} \sum_{\bf k} 
P (E_{\bf k})
\left[ 
E_{\bf k}^2 + T \xi_{\bf k} {\partial \mu \over \partial T} 
- {T \over 2} {\partial |{\bf d} ({\bf k})|^2 \over \partial T}
\right], 
$$
%%\end{equation}
%%
%
where 
$ P (E_{\bf k}) = 
f (E_{\bf k})
\left[ 1 -  f (E_{\bf k}) \right].
$
The results for $C_{\Gamma}$ can be seen in 
Fig.~\ref{fig:specific-heat}, and
analysed as follows.
The specific heat jump at $T_c = 1.5~{\rm K}$ is
characterized by the parameter
$\theta_{\Gamma} = C_{S, \Gamma} (T_c) / C_{N} (T_c) - 1$, 
which takes values
$\theta_{f_{xyz}} = 0.5596$, 
$\theta_{p_z} = 0.9733$, 
$\theta_{p_y} = 0.9358$, 
and 
$\theta_{p_x} = 1.2001$.
The corresponding singlet s-wave value
is $\theta_{s} = 1.4604$. 
An analysis of $C_{\Gamma}$ at low $T$ is 
also important in distinguishing
possible weak spin-orbit triplet phases.
Using the fact that $\partial \mu /\partial T \approx 0$,
$\partial |{\bf d} ({\bf k})|^2 /\partial T \approx 0$
at low $T$, then  $C_{\Gamma} \approx 2T \int_0^{\infty} dw w^2 
{\rm sech}^2 (w) N (2T w)$, where
$w \equiv \omega/2T$ and 
$N(\omega) = 2 \sum_{\bf k} \delta (\omega - E_{\bf k})$
is the auxiliary density of states.
This leads to 
$
C_{f_{xyz}} = (T/T_{f_{xyz}})^2 \log (K_{3} \Delta_{f_{xyz}}/T)
$,
where 
the temperature scale 
$
T_{f_{xyz}}^{-2} = 
\left[ 
2 
\left( 
\Delta_{f_{xyz}} t_x
\right)^{-1} 
D / \pi \kappa 
\right]
\gamma (3)
$,
with 
$
D = 
16/\left[ 
\pi |{\rm sin} (x)| 
\right]
$,
$
\gamma (m) = 
\int_{0}^{\infty} d w w^m {\rm sech}^2 (w),
$
$
\kappa = \sqrt {1 -(\mu/t_x)^2 } 
$
and 
$
K_{m} = 
\exp 
\left[ 
\int_0^\infty d w w^m {\rm sech}^2 (w) 
{\rm log} 
\left[ 
4 w^{-1} \kappa 
\right]
/ \gamma(m) 
\right].
$
This $T$ dependence of $C_{f_{xyz}}$ 
is a reflexion of the 
the behavior of $N(\omega) = \alpha (\omega/\Delta_0)
{\rm log} ( \beta \Delta_0/\omega )$ at 
small $\omega$, where $\alpha$ and $\beta$ are 
constants. The presence of the logarithmic term
results from the large number of low-energy 
states surrounding the double zeros of the
order parameter at the relatively flat 
Fermi surface. 
The $p_z$ and $p_y$ symmetries, however, have 
very different behaviors at low $T$, given
that both order parameters have lines of nodes
at the Fermi surface, but no double zeros.
In this case, the specific heats behave as 
$ C_{ p_{z} } = ( T/T_{p_z} )^2 $,
and $ C_{ p_{y} } = ( T/T_{p_y} )^2 $,
where 
$
T_{p_j}^{-2} = \left[ \Delta_{p_j} |t_x| \right]^{-1}
D \gamma(3),
$
for $ ( p_j = p_z~{\rm or}~p_y ) $. 
Lastly, 
$
C_{p_x} = 
(T_{p_x}/T)^{1/2}{\rm exp} (-\omega_0/T),
$
where
$\omega_0 = \Delta_{p_x} 
\sqrt{ 1 - 
\left[ 
\left(
|t_y| + |t_z| - \mu 
\right)
/ |t_x|
\right]^{2}
}
$
and
$T_{p_x}^{1/2} = A B \gamma (1/2)$,
with 
$
A = 2(\omega_0)^{7/2}
/
\left[
\pi^2  |{\rm sin} (x)| \Delta_{p_x}^2 (|t_y| + |t_z| - \mu)
\right]
$,
$
B = 
\left[ 
(t_x^2 - \Delta_{p_x}^2)/(t_y t_z)
\right]^{1/2}.
$
Notice that $C_{p_x}$ is exponentially
supressed at low $T$ due to the presence of a full gap in the
excitation spectrum, however the exponential prefactor behaves as
$T^{-1/2}$, which differs from the 
$T^{-3/2}$ behavior of the prefactor
in the singlet s-wave case. 
This difference arises because $N(\omega)$ in the $p_x$ case has a square root 
dependence in the vicinity of the gap edge,
$N_{p_x} \sim \sqrt { \omega - \omega_{0,p_x} } $,
while in the s-wave case $N(\omega)$ has a square root singularity
near the gap edge
$N_{s} \sim 1 / \sqrt { \omega - \omega_{0,s} } $.
The Gaussian correction $\Omega_G$ to $\Omega_0$ leads 
to collective modes at low $T$ with anisotropic dispersions $\omega = 
c_{\Gamma} (\theta,\phi) |{\bf q}|$ for all symmetries. These 
modes give a symmetry independent 
$T^2$ contribution to $C_{\Gamma}$,
but the prefactor is symmetry dependent. This collective
mode contribution is characteristic of neutral 
superfluids. However, in real charged superconductors that may become
plasmonized, and thus gapped.

\begin{figure}[ht!]
\begin{center}
\(
  \begin{array}{c}
   \multicolumn{1}{l}{\hspace{-0.2cm}\mbox{\bf (a)}} \\
	\psfrag{T (K)}{$T\ (K)$}
	\psfrag{Vpxx/a2 (K)}{$V_{0}\rho_{xx}/a_{x}^{2}\ (K)$}
     \includegraphics[width=7.0cm]{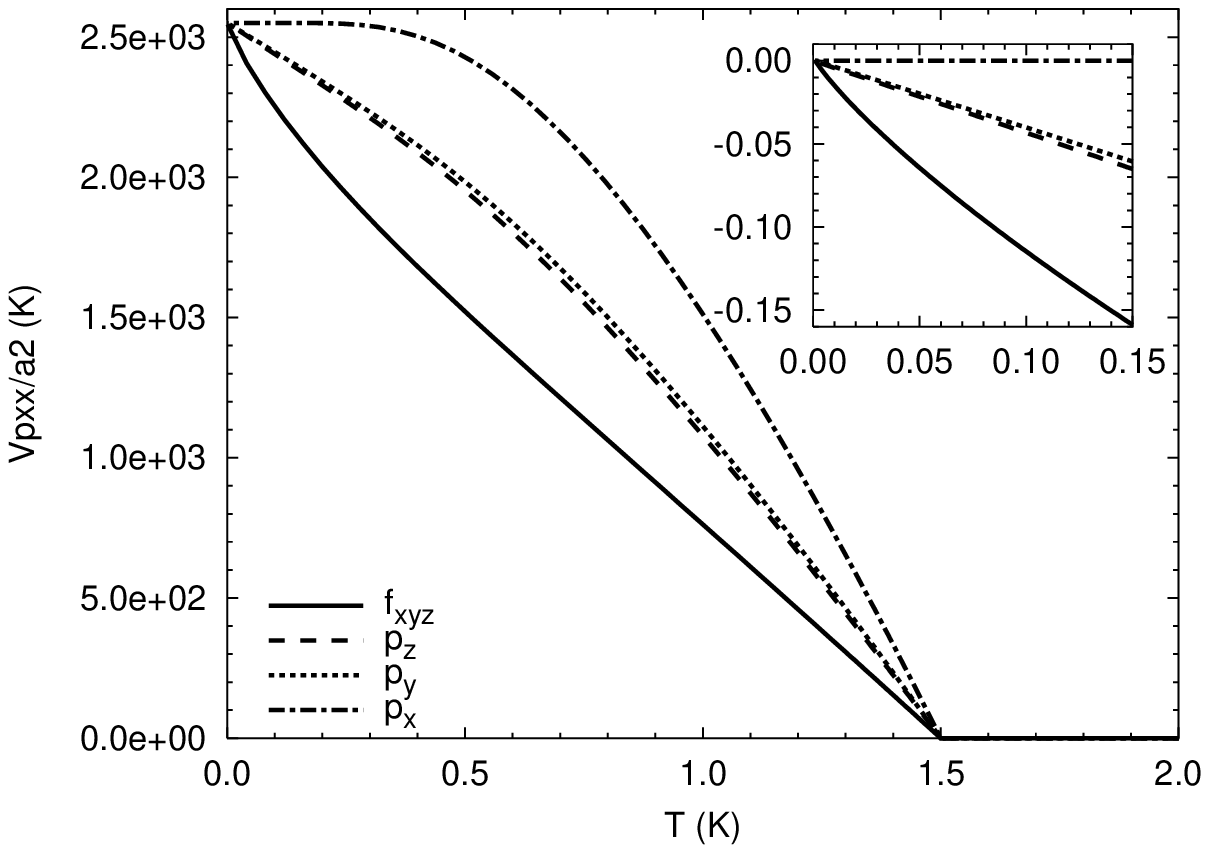} \\
   \multicolumn{1}{l}{\hspace{-0.2cm}\mbox{\bf (b)}} \\
     \psfrag{T (K)}{$T\ (K)$}
	\psfrag{Vpyy/b2 (K)}{$V_{0}\rho_{yy}/a_{y}^{2}\ (K)$}
\includegraphics[width=7.0cm]{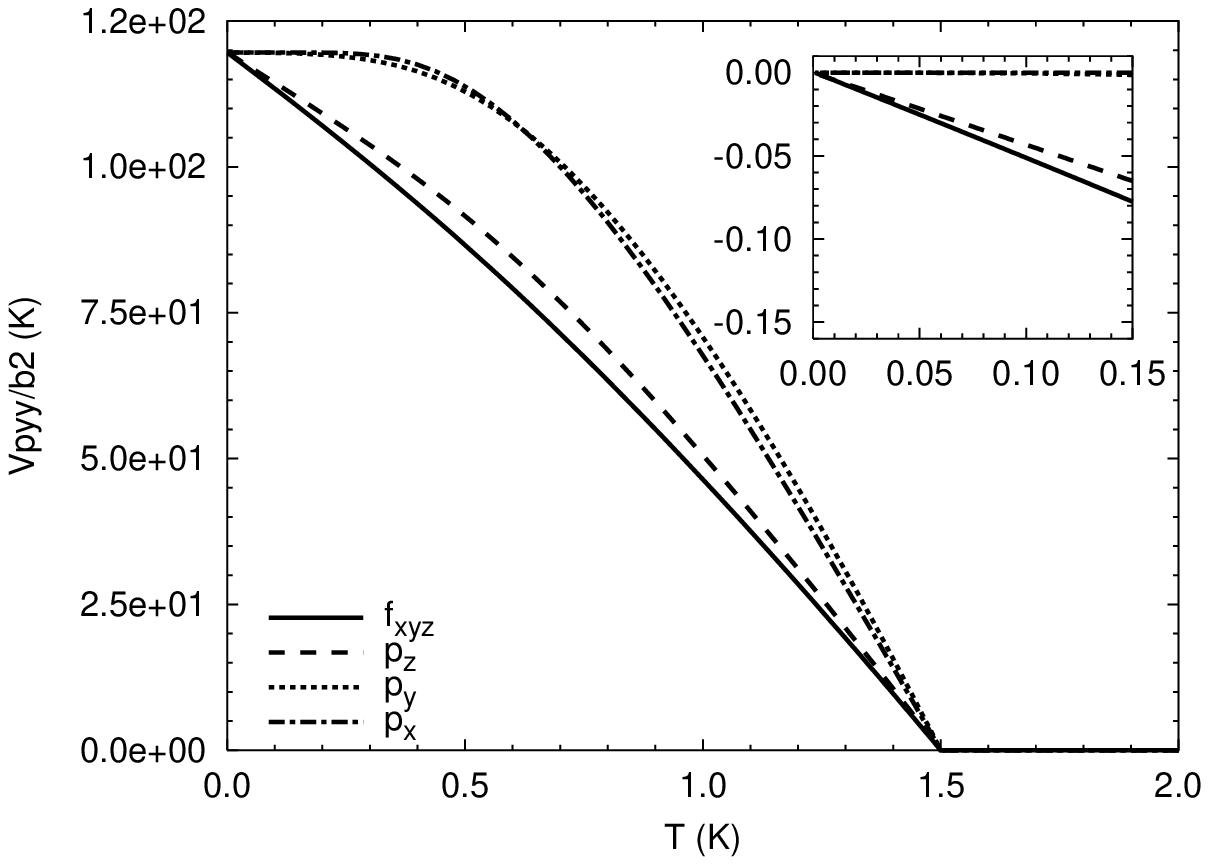} \\
   \multicolumn{1}{l}{\hspace{-0.2cm}\mbox{\bf (c)}} \\
     \psfrag{T (K)}{$T\ (K)$}
	\psfrag{Vpzz/c2 (K)}{$V_{0}\rho_{zz}/a_{z}^{2}\ (K)$}
\includegraphics[width=7.0cm]{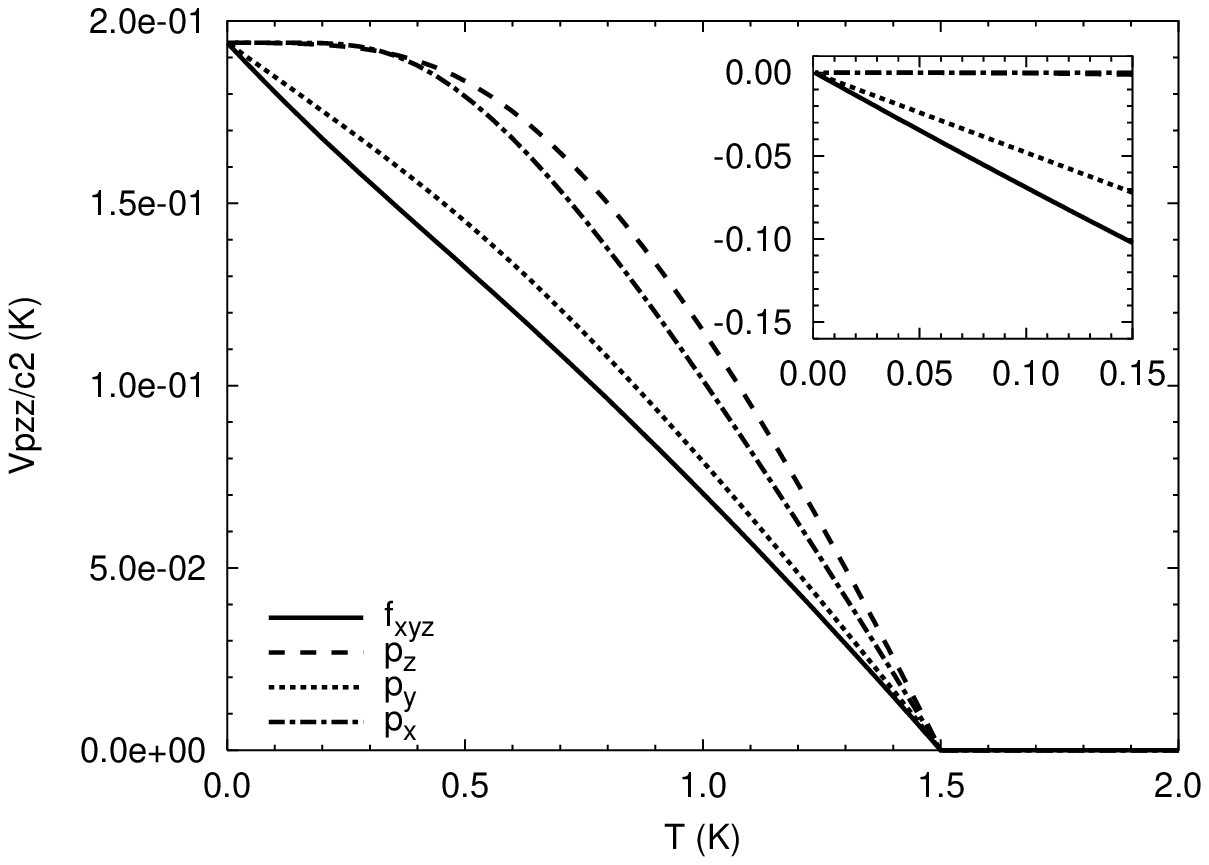} 
\end{array}\)
\end{center}
\caption{
Temperature dependence of $V_0 \rho_{ii}/a_i^2$ in thermal 
units $(K)$.
$V_0$ $(a_i)$ is the unit cell volume (length).
Insets show the low $T$ behavior of 
$
\Delta \rho_{ii} 
\equiv 
\left[
\rho_{ii} (T)/\rho_{ii} (0) - 1 \right
].
$
}
\label{fig:superfluid-density}
\end{figure}

The temperature dependence of the superfluid density tensor $\rho_{ij} (T)$
can also be used to distinguish different weak spin-orbit 
triplet phases. This tensor is directly associated with phase
twists of the $U(1)$ phase of the ${\bf d}$-vector.
Take 
$
{\bf d} ({\bf k}) \to {\bf d} ({\bf k}) 
{\rm exp} \left[ i\phi ({\bf k}) \right]
$
and 
expand 
$
S_{\rm eff}
$
in powers of $\phi ({\bf k})$
about the saddle point with 
$\phi ({\bf k}) = 0$.
The resulting action 
$\Delta S = S_{\rm eff} (\phi) - S_{\rm eff} (\phi = 0)$ 
is
$
\Delta S = - {V/2} \sum_{\bf} 
\phi ({\bf k}) \phi ({\bf - k}) k_i k_j \rho_{ij},
$
with
\begin{equation}
\label{eqn:superfluid-density}
\rho_{ij} (T) = {1 \over V} 
\sum_{\bf k}
\left[ n_{\bf k}
{\partial_i} {\partial_j} \xi_{\bf k}
-
Y_{\bf k}
{\partial_i} \xi_{\bf k} {\partial_j} \xi_{\bf k} 
\right],
\end{equation}
where $n_{\bf k}$
is the momentum distribution, and 
$
Y_{\bf k} = (2T)^{-1} 
{\rm sech}^2 \left( { E_{\bf k} / 2 T} \right)
$
is the Yoshida distribution.
In Fig.~\ref{fig:superfluid-density} we show
the $T$ dependence of $\rho_{ij}$. 
In the case of the $D_{2h}$ group only diagonal components
$\rho_{ii}$ exist, but they are highly 
anisotropic due to the quasi-one-dimensionality
of $\xi_{\bf k}$. 
The low $T$ behavior of $\Delta \rho_{ii} \equiv  
\left[ \rho_{ii} (T) / \rho_{ii} (0) - 1\right]$
is shown in the inset of Fig.~\ref{fig:superfluid-density}.
At low T, the main contributions to $\Delta \rho_{ii}$
come from the term containing $Y_{\bf k}$.
For the $p_x$ symmetry,
$\Delta \rho_{xx} = 
- (T/T^{x}_{p_x})^{1/2} {\rm exp} (-\omega_0/T)$,
$\Delta \rho_{yy} = 
- (T/T^{y}_{p_x})^{3/2} {\rm exp} (-\omega_0/T)$,
$\Delta \rho_{zz} = 
- (T/T^{z}_{p_x})^{3/2} {\rm exp} (-\omega_0/T)$,
where $T^{i}_{p_x}$ are characteristic temperatures.
Notice the exponential behavior due to presence of a full gap. 
Furthermore, notice that the $T$ dependence of the prefactor
of $\Delta \rho_{xx}$ is different from $\Delta \rho_{yy}$ 
and $\Delta \rho_{zz}$ due to the highly anisotropic Fermi 
surface of these systems, i.e., the velocity $v_x = \partial_{x} \xi_{\bf k}$ 
does not vanish anywhere at the Fermi surface.
For the $p_y$ symmetry,
$\Delta \rho_{xx} = - T/T^{x}_{p_y}$,
$\Delta \rho_{yy} = - (T/T^{y}_{p_y})^{3}$, and 
$\Delta \rho_{zz} = - T/T^{z}_{p_y}$.
Notice the $T^3$ power law for $\Delta \rho_{yy}$, 
which results from the simultaneous contribution 
from the lines of nodes of ${\bf d}_{p_y}$ and 
the zeros of $v_y = \partial_y \xi_{\bf k}$.  
Similar behavior is found also 
for the $p_z$ symmetry,
$\Delta \rho_{xx} = - (T/T^{x}_{p_z})$,
$\Delta \rho_{yy} = - (T/T^{y}_{p_z})$, and
$\Delta \rho_{zz} = - (T/T^{z}_{p_z})^{3}$.
In the $p_z$ case, however, the $T^3$ dependence appears in 
$\Delta \rho_{zz}$.
Finally for the $f_{xyz}$ symmetry,
$\Delta \rho_{xx} = - (T/T^{x}_{f_{xyz}}) {\rm log} (K_{1} \Delta_{f_{xyz}}/T)$
$\Delta \rho_{yy} = - T/T^{y}_{f_{xyz}}$, and 
$\Delta \rho_{zz} = - T/T^{z}_{f_{xyz}}$.
The logarithmic dependence on  
$\Delta \rho_{xx}$ originates from  the node structure of
${\bf d}_{f_{xyz}}$ and from a non-vanishing $v_x$ at the Fermi surface.
Notice that the logarithmic dependence is absent on
$\Delta \rho_{yy}$ and $\Delta \rho_{zz}$, because the zeros
of $v_y$ and $v_z$ cancel it out. The $T$
dependence of $\rho_{ii} (T)$ can be measured via penetration
depth experiments.

In summary, we studied several properties
of organic quasi-one-dimensional conductors (Bechgaard salts) 
which are strong
candidates for triplet superconductivity.
We compared weak spin-orbit coupling symmetries 
$f_{xyz}$, $p_x$, $p_y$ and $p_z$ via several
thermodynamic quantities. For each symmetry, we analysed the 
temperature dependences of the order parameter, 
condensation energy, specific heat, and superfluid density tensor.
We would like to thank NSF (Grant No. DMR-9803111) for support. 
$^\dagger$Present address: Department of Physics, 
Massachusetts Institute of Technology,
Cambridge, MA 02139-4307.

%\bibliography{fwave-pwave}% Produces the bibliography via BibTeX.

\end{document}